\documentstyle[prl,aps,epsfig,twocolumn]{revtex}

\begin{document}
\draft

\twocolumn[\hsize\textwidth\columnwidth\hsize\csname@twocolumnfalse\endcsname
\title{Geometric Quantum Information Storage Based on Atomic Ensemble}
\author{C. P. Sun$^{1,2,a,b}$, P. Zhang$^{1}$ and Y. Li$^{1}$}
\address{$^{1}$Institute of Theoretical Physics, the Chinese Academy of Science,
 Beijing, 100080, China}
 \address{$^{2}$Department of Physics, Nankai University, Tianjin 300070, China}
\maketitle

\begin{abstract}
Quantum information storage (QIS) is a physical process to write
quantum states into a quantum memory (QM). We observe that in some
general cases the quantum state can be retrieved up to a unitary
transformation depicted by the non-Abelian Berry's geometric phase
factor (holonomy). The QIS of photon with this geometric character
can be implemented with the symmetric collective excitations of a
$\Lambda$-atom ensemble, which is adiabatically controlled by a
classical light with a small detuning $\delta$. The cyclic change of
the controlled Rabi frequency $\Omega(t)$ with period
$T=\frac{2n\pi}{\delta}$ accumulates additional geometric phases in
different components of the final photon state. Then, in a purely
geometric way the stored state can be decoded from the final state.
\end{abstract}
 ]

An efficient quantum information storage (QIS) requires a quantum
memory (QM), which is a robust information recording system with a
decoherence time scale much larger than the time scale of information
processing \cite{q-infor}. With these two time scales the QIS can be
understood as a controlled dynamic process transferring the
information encoded in quantum states into the QM. When one does not
process the transferred information in QM to test the efficiency of
the QM, the readout in QIS is implemented by the dynamics retrieving
the input quantum data in the output. Thus the identity of input and
output implies a cyclic evolution in some sense. On the other hand,
it is well known that the Berry geometric phase factor (BGPF) can
occur in an adiabatically cyclic evolution \cite{BPF}. This
intuitional consideration motivates us to probe the geometric effects
in the QIS.

In this article, we take as an example the atomic ensemble based
quantum memory \cite{Lukin1,Lukin2,Lukin3} associated with the
electromagnetically induced transparency (EIT) effect \cite{EIT} to
show that, in the degenerate case, the generalization of the idea of
the BGPF (or called the non-Abelian Berry's holonomy) \cite{ZW} might
make possible in a quantum state storage a purely geometric decoding
way, which depends on neither the state to be stored, nor the dynamic
details controlling the interaction. Actually, the BGPF and its
generalizations have attracted considerable attention in both
experimental and theoretical aspects of quantum computing
$\cite{Zana,Ekert,Duan}$. People can take advantage of this geometric
features to the aim of quantum information processing, as the
robustness of the BGPF can result in a resilience against some kinds
of decoherence sources $\cite{Fazio,DU,WZD}$.

The original scheme of atomic ensemble based QIS requires both the
quantum probe light and the classical control laser to be tuned onto
resonance with the two Rabi transitions of $\Lambda $-atom
respectively \cite{Lukin1,Lukin2,Lukin3}. A most recent protocol for
QIS based on the quasi-spin wave excitation in an "atomic crystal"
\cite{Sun-prl} instead of the atomic vapor ensemble also requires the
same resonance condition to avoid the spatial-motion induced
decoherence \cite{Sun-pra}. However, in this article we can prove
that, even when the classical control laser is not on resonance, the
atomic ensemble based QIS can still work well. Here the crucial point
is the BGPF occurring in the retrieved output states.

We first describe the general idea of QIS associated with the
non-Abelian Berry's phase factor (NABPF). Let $M$ be a QM possessing
a subspace spanned by $|M_{n}\rangle ,n=1,2,...,d$, which can store
the quantum information of a system $S$ with the basis vectors
$|s_{n}\rangle ,n=1,2,...,d$. The controlled interaction Hamiltonian
$H=H(t)=H[{\bf R(t)}]$ of period $T$ depends on
time through the adiabatically varying parameters $%
{\bf R(t)}=(R_{1}(t),R_{2}(t),...,R_{s}(t))$. If there exists a set
of degenerate ground states $|E_{n}(t)\rangle $ of $H$ interpolating
between the initial state $|s_{n}\rangle $ $\otimes |M\rangle $ and
the final state $|s\rangle \otimes |M_{n}\rangle $ for each index $n$
and the given states $|s\rangle $ and $|M\rangle $, we can define the
geometric QIS when the NABPF $W(t)$ has a factorization structure.
Mathematically, $|E_{n}(t)\rangle $ and $W(t)$ are required to
satisfy
\begin{eqnarray}
|E_{n}(0)\rangle  &=&|s_{n}\rangle \otimes |M\rangle
,|E_{n}(T_{m})\rangle
=|s\rangle \otimes |M_{n}\rangle,   \nonumber \\
|E_{n}(T)\rangle  &=&|s_{n}\rangle \otimes |M\rangle,  \\
W(T_{m}) &=&1\otimes W_{M},W(T)=W_{S}\otimes 1  \nonumber
\end{eqnarray}%
at two instances $T_{m}(<T)$ and $T$. In fact, writing an arbitrary
state $|s(0)\rangle =\sum_{n}c_{n}\,|s_{n}\rangle $ of $S$ into $M$
with the initial state $|M\rangle $ can be realized as a controlled
evolution from time $t=0$ to $t=T_{m}$
\begin{equation}
\sum_{n}c_{n}|s_{n}\rangle \otimes |M\rangle \rightarrow |s\rangle \otimes
W_{M}\sum_{n}c_{n}|M_{n}\rangle
\end{equation}%
The readout process from $M$ is another controlled evolution from time $%
t=T_{m}$ to $t=T$
\begin{equation}
|s\rangle \otimes W_{M}\sum_{n}c_{n}|M_{n}\rangle \rightarrow
(W_{S}\sum_{n}c_{n}|s_{n}\rangle )\otimes |M\rangle
\end{equation}%
These two processes combine into a cyclic evolution with the NABPF.
Because the factor $W_{S}=W_{S}[C]$ is known to be geometric
$|s_{n}\rangle $ can be easily decoded from $|f_{n}\rangle
=W_{S}|s_{n}\rangle $ in a purely-geometric way.

\begin{figure}[h]
\includegraphics[width=6cm,height=6cm]{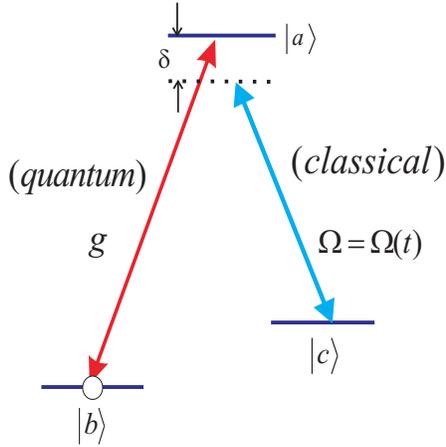}
\caption{A 3-level atom with off-resonance EIT: The probe light is
resonantly coupled to atomic transition while the control light field
is coupled to the atomic transitions with the same detuning
$\protect\delta$. }
\end{figure}

Now, we consider an atomic ensemble based implementation of the above
protocol of geometric QIS for photon. Our ensemble consists of $N$
identical 3-level system of $\Lambda $-type, which are coupled to two
single-mode optical fields as shown in Fig. 1. The atomic levels are
labelled as the ground state $|b\rangle $, the excited state
$|a\rangle $ and the meta-stable state $|c\rangle $. The atomic
transition $|a\rangle \rightarrow |b\rangle $ with energy level
difference $\omega _{ab}$ $=\omega _{a}-\omega _{b}$ is resonantly
coupled to a quantized probe light of frequency $\omega = $ $\omega
_{ab}$ with the coupling coefficient $g$; and the atomic transition
$|a\rangle \rightarrow |c\rangle $ with energy level difference
$\omega _{ac}$ is driven by a classical control field of frequency
$v\neq \omega _{ac}$ with the Rabi-frequency $\Omega (t)$. We suppose
that there is a tiny detuning $\delta $ $=\omega _{ac}-v$. Let $\hbar
=1$. Then under the
rotating wave approximation the interaction Hamiltonian can be written as%
\cite{Sun-prl}

\begin{equation}
H_{I}=g\sqrt{N}aA^{\dagger }+\Omega \exp [i\phi (t)]S_{+}+h.c.,
\end{equation}%
in terms of the symmetrized collective atomic operators
\begin{eqnarray}
A &=&\frac{1}{\sqrt{N}}\sum_{j=1}^{N}\sigma _{ba}^{j},  \nonumber \\
S_{-} &=&\sum_{j=1}^{N}\sigma _{ca}^{j},\text{
}S_{+}=(S_{-})^{\dagger },
\end{eqnarray}%
where $\sigma _{\mu \nu }^{j}=|\mu \rangle _{jj}\langle \nu |$ is the flip
operator of the $j^{th}$ atom from states $|\mu \rangle _{j}$ to $|\nu
\rangle _{j}$ $(\mu ,\nu =a,b,c)$; and $a^{\dagger }$ and $a$ the creation
and annihilation operators of quantized probe light respectively. The
coupling coefficients $g$ and $\Omega $ are real and assumed to be identical
for different atoms in the ensemble. Most recently a similar effective
Hamiltonian was given by us for the case of atomic crystal \cite{Sun-prl},
with the spin-wave type collective atomic operators replacing the above
symmetrized ones.

The slight difference between the quantum memory models of the
on-resonance EIT and the off-resonance EIT lies in the slowly
changing time-dependent phase $\phi (t)=\delta t$. Actually the
on-resonance EIT is not a prerequisite for achieving significant
group velocity reduction \cite{Lukin2} and thus the atomic ensemble
with off-resonance EIT can also be recognized as a QM. More
generally, the EIT phenomenon can occur even when the frequency
difference between the probe and control lasers matches the two-photo
transition between the two lower states of the $\Lambda $-type atoms.
The theoretical and experimental results on significant
group-velocity reduction have been reported in this case
\cite{Deng1,Deng2}.

Let us first consider dynamic symmetry in the low excitation of
atomic ensemble with most of $N$ atoms staying in the ground state
$|b\rangle $ and
$N\rightarrow \infty $, which was first discovered most recently by us \cite%
{Sun-prl} for the quasi-spin wave case. Together with the third generator $%
S_{3}=\sum_{j=1}^{N}(\sigma _{aa}^{j}-\sigma _{cc}^{j})/2$, $S_{-}$ and $%
S_{+}$ generate the $SU(2)$ algebra. The commutation relation $%
[A,S_{+}]=C$ of $A$ with the $SU(2)$ generators defines another
collective operator
\begin{equation}
C=\frac{1}{\sqrt{N}}\sum_{j=1}^{N}\sigma _{bc}^{j},
\end{equation}
satisfying the commutation relation $[C,S_{-}]=A$. As a special case of
quasi-spin wave excitation with zero varying phases, the above two mode
symmetrized excitations defined by $A$ and $C$ behave as two independent
bosons since they satisfy $[A,A^{\dagger }]=1$ and $[C,C^{\dagger }]=1$ in
the large $N$ limit with the low excitation \cite{Sun-prb}. The commutation relation $%
[SU(2),h^{2}]\subset h^{2}$ of commutators between $SU(2)$ algebra and the
Heisenberg-Weyl algebra $h^{2}$ generated by $A$, $A^{\dagger }$,$C$ and $%
C^{\dagger }$  implies that the dynamic symmetry of evolution governed by $H$
can be depicted by the semi-direct product algebra $SU(2)\overline{\otimes }%
h^{2}$.

Taking the additional phase factor $\exp [i\phi ]$ in the above effective
Hamiltonian into account, we can write down the dark(bright)-state polariton
(DSP(BSP)) operator
\begin{eqnarray}
D &=&D(\theta ,\phi )=a\cdot \cos \theta -C\cdot \sin \theta e^{i\phi
},
\nonumber \\
B &=&B(\theta ,\phi )=a\cdot \sin \theta +C\cdot \cos \theta e^{i\phi }
\end{eqnarray}%
satisfying the bosonic commutation relations. Here, the extent of mixing of
photonic and atomic degrees of freedom is controlled by the Rabi frequency $%
\Omega (t)$ of the atomic transition $|a\rangle \rightarrow |c\rangle
$, through the relation
\begin{equation}
\tan \theta =g\sqrt{N}/\Omega (t).
\end{equation}%
It is easy to check that $[H_{I},D]=0$.  Thus we readly obtain a family of
instantaneous dark-states, i.e. the eigen-states of $H_{I}(t)$ with
vanishing eigenvalues
\begin{equation}
|d_{n}(t)\rangle =|d_{n}[\theta ,\phi ]\rangle =\frac{1}{\sqrt{n!}}%
D^{\dagger n}|{\bf 0}\rangle \text{ }
\end{equation}%
where $|{\bf 0}\rangle =|0\rangle \otimes |{\bf b}\rangle \equiv |0\rangle
\otimes |b,b,...,b\rangle $ represents the ground state of the total coupled
system with each atom in the ground state $|b\rangle $ and the light field
in the vacuum state $|0\rangle $. The degenerate dark states $%
|d_{n}(t)\rangle $ $(n=0,1,2,...)$ span the storage space ${\bf
V}(t)$.

Now, let us consider the geometric QIS implemented by DSP based on
the above zero eigenvalue state $|d_{n}(t)\rangle $. Quantum
information is initially encoded as a superposition of photonic
states $|s(0)\rangle =\sum_{n}c_{n}\,|n\rangle $ while all atoms are
prepared in the ground state. Then the total initial state is $|\Phi
(0)\rangle =\sum_{n}c_{n}\,|n\rangle \otimes |{\bf b}\rangle =$
$\sum_{n}c_{n}|d_{n}(0) \rangle \in $ $V(t)$ for very large $\Omega
(0)$ and $g\sqrt{N}/\Omega (0)\rightarrow 0 $. It follows from the
quantum adiabatic theorem for degenerate case \cite{zee,prd} that
under the adiabatic conditions
\begin{equation}
\frac{g\sqrt{N}\stackrel{.}{|\Omega |}}{(\sqrt{g^{2}N+\Omega ^{2}})^{3}},%
\frac{g\sqrt{N}\left\vert \delta \Omega \right\vert }{(\sqrt{g^{2}N+\Omega
^{2}})^{3}}\ll 1,
\end{equation}%
the adiabatic evolution of the degenerate system will keep itself
within the block spanned by those degenerate eigen-states with the
same instantaneous eigen-value $0$. This adiabatic condition requires
that the detuning be small enough to forbid the transitions among
$|d_{n}(t)\rangle (n=0,1,2,...) $. In this circumstance, the
corresponding wave function $|\Phi (t)\rangle
=\sum_{n}c_{n}(t)|d_{n}(t)\rangle $ is governed by the matrix
equation \cite{zee,prd}
\begin{equation}
\frac{d}{dt}{\bf C}(t)=K(t){\bf C}(t),
\end{equation}%
where the vector ${\bf C}(t)$ of coefficients and the connection matrix $K(t)
$ are respectively defined by
\begin{eqnarray}
{\bf C}(t) &=&(c_{0}(t),c_{1}(t),c_{2}(t),...)^{T},  \nonumber \\
K(t) &=&(-\langle d_{m}(t)|\partial _{t}d_{n}(t)\rangle
)_{m,n=0,1,2,...}.
\end{eqnarray}%
The solution ${\bf C}(t)=W(t)$ ${\bf C}(0)$ formally determines a
time-ordered integral $W(t)=T\exp \{\int K(t)dt\}$ as the NABPF.

The non-diagonalized NABPF can mix different instantaneous
eigenstates $|d_{n}(t)\rangle $ to induce a non-Abelian gauge
structure. Thus in general it is very difficult to obtain the
explicit expression for the NABPF. Fortunately, the above NABPF
$W(t)$ in our protocol of QIS can be Abelianized as a diagonal
matrix since we can prove that $K(t)$ is diagonal. In fact, by
straightforward calculation we can address the adiabatic motion
equations for DSP and BSP:
\begin{eqnarray}
\stackrel{\cdot }{D} &=&-(\stackrel{\cdot }{\theta }+i\delta \sin \theta
\cos \theta )\cdot B+i\delta \sin ^{2}\theta \cdot D,  \nonumber \\
\stackrel{\cdot }{B} &=&(\stackrel{\cdot }{\theta }-i\delta \sin
\theta \cos \theta )\cdot D+i\delta \cos ^{2}\theta \cdot B.
\end{eqnarray}
It is then easy to prove that $K(t)_{m,n}\propto \delta _{mn}$ and
thus the NABPF reduces to a diagonal matrix with the elements
$W(t)_{mn}=\delta _{mn}\exp \{i\gamma _{n}(t)\}$  where the
Abelianized Berry phase is
\begin{equation}
\gamma _{n}(t)=i\int_{0}^{t}\langle d_{n}(\tau )|\frac{d}{d\tau }|d_{n}(\tau
)\rangle d\tau \equiv n\gamma (t)
\end{equation}%
with $\gamma (t)=-\int_{0}^{t}\stackrel{\cdot }{\phi }\sin ^{2}\theta d\tau $%
..

Now with the above expression we can answer the question: what is the
geometric nature of this Berry phase for the QIS? The geometric
nature can readily be seen by introducing the equivalent parameter
space spanned by
\begin{equation}
{\bf R}{\bf =(}R_{1},R_{2},R_{3})=(g\sqrt{N}\cos \phi ,g\sqrt{N}\sin \phi
,\Omega (t))
\end{equation}%
where $R=\sqrt{g^{2}N+\Omega ^{2}}$. When $\Omega (t)$ has the same period $%
T=\frac{2\pi }{\delta }$ as that of $\phi $, the path ${\bf R(t)}$ can be
imagined as a contour ${\bf C}$ on the cylinder defined by $%
R_{1}^{2}+R_{2}^{2}=g^{2}N$, one point of which is attached on the $%
R_{1}-R_{2}$ plane with $R_{3}=0$ (see Fig. 2). Here, we have chosen
the
coordinate origin coincident with ${\bf R=0}$ for the centrifugal vector $%
{\bf R}$. Then the dynamic process of QIS with the BGPF implements as the
parameter $\Omega (t)$ is adiabatically manipulated.

\begin{figure}[h]
\includegraphics[width=5cm,height=6cm]{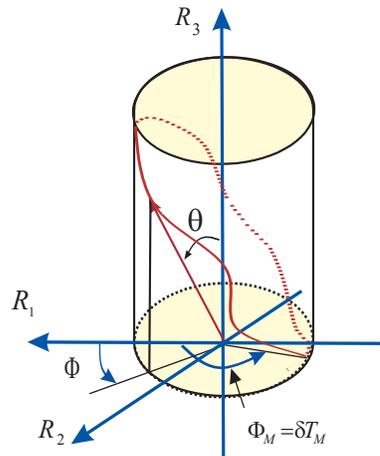}
\caption{The topology of the QIS: a closed path in parameter space is
attached on the cylinder by $R_{1}^{2}+R_{2}^{2}=g^{2}N$. The writing
process is realized by changing the point along the path from for a
very large value $\Omega (t)$ at $t=0$ to
$(g\sqrt{N}\cos\phi_M,g\sqrt{N}\sin\phi_M, 0)$  at $t=T_{M}$; while
the readout process is realized by changing the point along the path
from $(g\sqrt{N}\cos\phi_M,g\sqrt{N}\sin\phi_M, 0)$ to $(g\sqrt{N},0,
\Omega (t) )$ inversely.}
\end{figure}

As illustrated in Fig. 2, to write the information carried by the
photonic state $|\varphi \rangle =\sum_{n}c_{n}\,|n\rangle $ into the
atomic ensemble based memory, we change $\Omega (t)$ from a very
large value in comparison with $g\sqrt{N}$ at $t=0$ to zero at
$t=T_{M}$ $(0<T_{M}<T)$ so that the phase $\theta $ changes from $0$
to $\frac{\pi }{2}$. Correspondingly, the
initial state $|\Phi (0)\rangle =\sum_{n}c_{n}\,|n\rangle \otimes |{\bf b}%
\rangle $ $\ $ evolves into the state of excitonic type

\begin{eqnarray}
|\Phi (t=T_{M})\rangle &=&\sum_{n}c_{n}|d_{n}(T_{M})\rangle   \nonumber \\
&=&|0\rangle \otimes \sum_{n}\exp [in\gamma
(T_{M})](-1)^{n}c_{n}|n\rangle _{c},
\end{eqnarray}%
where, $|n\rangle _{c}=(1/\sqrt{n!})C^{\dagger n}|{\bf b}\rangle $ is
a many-atom Dicke state \cite{Dicke}. The read-out process with an
information processing, which is quantum mechanically a unitary
transformation $U_{p}(t)$, can be realized by the manipulation of
$\Omega (t)$ from $0$ to a very large value
as $\theta $ changes from $0$ to $\frac{\pi }{2}$ at the instance $T=\frac{%
2\pi }{\delta }$. To test the efficiency of the quantum memory, we can test
the output without the middle information processing
\begin{equation}
|\Phi (T)\rangle =(\sum_{n}c_{n}\exp [in\gamma _{C}]|n\rangle )\otimes |{\bf %
b}\rangle.
\end{equation}%
The result of manipulating $\Omega (t)$ from $\infty $ to $0$ and then from $%
0$ to $\infty $ inversely is to multiply the Fock state of probe photon by
the Berry topological phases $\gamma _{n}[C]=n\gamma \lbrack C]:$
\begin{equation}
\gamma \lbrack C]=-\int_{0}^{2\pi }\sin ^{2}\theta d\phi =\int_{0}^{2\pi }%
\frac{-g^{2}Nd\phi }{g^{2}N+F(\phi )^{2}},
\end{equation}%
where $F(\phi )=\Omega (\phi /\delta )$ satisfies the "fixed point
conditions"
\begin{equation}
F(0)\rightarrow \infty ,F(\phi _{M})\rightarrow 0,F(2\pi )\rightarrow
\infty,
\end{equation}%
where $\phi _{M}=\delta T_{M}$. The topological properties of $\gamma
\lbrack C]$ is reflected by the fact that the integrals along
different paths on the cylinder through the two fixed points
$(g\sqrt{N},0,\infty )$ and $(0,g\sqrt{N},0)$ can span the same solid
angle and thus lead to the same BGPF. The cyclic evolution combined
by the writing and readout processes is accompanied by the factorized
NABPF
\begin{equation}
W_{S}(T)=\sum_{n}\exp (in\gamma \lbrack C])|n\rangle
_{cc}\langle n|\otimes I
\end{equation}
which is known to be geometric. Therefore, $|\varphi \rangle
=\sum_{n}c_{n}\,|n\rangle _{c}$ can be easily decoded from
\begin{equation}
|f_{n}\rangle =W_{S}|\varphi \rangle =\sum_{n}c_{n}\exp (in\gamma \lbrack
C])|n\rangle _{c}
\end{equation}%
in a purely-geometric way.

Before concluding this article it is worthy to point out that, just
as the system of quasi-spin wave collective excitations proposed most
recently by us \cite{Sun-prl}, the present atomic ensemble based QM
also possesses an instantaneous hidden dynamic symmetry in the large
$N$ limit with low excitations and thus the extra degenerate
eigen-states besides the collective dark states. However, we also
prove that, under the adiabatic conditions, there is no mixing among
the dark states and the extra degenerate eigen-states. That is to
say, the adiabatic change of external parameters does not lead the
system to enter into the subspace of these extra degenerate
eigen-states from our memory space. These theoretical results are
crucial for the protocol of implementing the geometric quantum
memory.

In conclusion, we have presented a generalized version of QIS by
allowing the quantum state to be retrieved up to an input-independent
unitary transformation depicted by the NABPF. Thus, to decode the
ideal input state, we only need to consider the geometry of the
parameter space determined by the change of parameters. This novel
geometric property of the QM and the corresponding QIS is independent
of both the state to be stored and the dynamic details of interaction
control. We also demonstrate the physical process of the geometric
QIS of photon by means of the symmetric collective excitations of a
$\Lambda $-atom ensemble adiabatically controlled by an off-resonance
classical light with a small detuning.

We acknowledge the support of the CNSF (grant No. 90203018) and the
Knowledged Innovation Program (KIP) of the Chinese Academy of
Sciences and the National Fundamental Research Program of China with
No. 001GB309310. We also sincerely thank L. You and X. F. Liu for
helpful discussions.

\end{document}